\documentclass{aa}
\usepackage{graphicx}
\begin{document}
   \title{Grand minima and maxima of solar activity: New observational constraints}


   \author{I.G. Usoskin\inst{1}, S.K. Solanki\inst{2},
          \and
          G.A. Kovaltsov\inst{3}
          }

   \offprints{I.G. Usoskin}

   \institute{Sodankyl\"a Geophysical Observatory (Oulu unit),
POB 3000, University of Oulu, Finland\\
\email{ilya.usoskin@oulu.fi}
         \and
             Max-Planck-Institut f\"ur Sonnensystemforschung, 37191
Katlenburg-Lindau, Germany
   \and
    Ioffe Physical-Technical Institute, Politekhnicheskaya 26,
RU-194021 St. Petersburg, Russia
             }

   \date{Received Month XX, 2007; accepted Month XX, 2007}


  \abstract
   {}
   {Using a reconstruction of sunspot numbers stretching over multiple millennia,
     we analyze the statistics of the occurrence of grand minima and maxima
     and set new observational constraints on long-term solar and stellar dynamo models.
     }
   {We present an updated reconstruction of sunspot number over multiple millennia,
    from $^{14}$C data by means of a physics-based model, using an updated model of the evolution of
    the solar open magnetic flux.
    A list of grand minima and maxima of solar activity is presented for
     the Holocene (since 9500 BC) and the statistics of both the length of individual
     events as well as the waiting time between them are analyzed.
   }
   {The occurrence of grand minima/maxima is driven not by long-term cyclic
      variability, but by a stochastic/chaotic process.
The waiting time distribution of the occurrence of grand minima/maxima deviates from
 an exponential distribution, implying that these events tend to cluster
 together with long event-free periods between the clusters.
Two different types of grand minima are observed: short (30--90 years) minima of Maunder
 type and long ($>$110 years) minima of Sp\"orer type, implying that a
 deterministic behaviour of the dynamo during a grand minimum defines its length.
The duration of grand maxima follows an exponential distribution, suggesting
 that the duration of a grand maximum is determined by a random process.
}
{These results set new observational constraints upon the long-term
 behaviour of the solar dynamo.
 }

\authorrunning{Usoskin et al.}
\titlerunning{Grand minima/maxima of solar activity}

   \maketitle
%

\section{Introduction}

The Sun is the only star whose magnetic activity can be studied on long time scales.
Direct solar observations since 1610 reveal great variability of the cycle averaged
 magnetic activity level of the Sun
 -- from the extremely quiet Maunder minimum (second half of 17th century) up to the modern
 episode of enhanced activity since the middle of the 20th century.
The Maunder minimum is representative of grand minima of solar activity (e.g., \cite{eddy77}),
 when sunspots almost completely vanished from the solar surface, while the solar wind
 appeared to continue blowing, although at a reduced pace (\cite{cliv98,usos01}).
A grand minimum is believed to correspond to a special state of the dynamo
 (\cite{soko04,miya06}), and its very existence poses a challenge for solar dynamo theory.
It is noteworthy that dynamo models do not agree how often such episodes occur in the Sun's history
 and whether their appearance is regular or random.
For example, the commonly used mean-field dynamo yields a fairly regular 11-year cycle,
 while there are also dynamo models including a stochastic driver (e.g., \cite{chou92,schm96,osse00, weis00,mini01,char01}, 2004)
 which predict intermittency of the solar magnetic activity.
The presence of grand maxima of solar activity has been mentioned (\cite{eddy77,usos03,sola04})
 but has not yet been studied in great detail.

Thanks to the recent development of precise technologies, including accelerator mass
 spectrometry, solar activity can be reconstructed over multiple millenia
 from concentrations of cosmogenic isotopes $^{14}$C and $^{10}$Be in terrestrial archives.
This allows one to study the temporal evolution of solar magnetic activity,
 and thus of the solar dynamo, on much longer time scales than available from
 direct measurements.
Consequently, a number of attempts to investigate the occurrence of grand minima in the past,
 using radiocarbon $^{14}$C data in tree rings, have been undertaken.
E.g., Eddy (1977a) identified major excursions in the available $^{14}$C record as
 grand minima and maxima of solar activity and presented a list of 6 grand minima
 and 5 grand maxima for the last 5000 years.
Stuiver \& Braziunas (1989) studied grand minima as systematic excesses of the high-pass
 filtered $^{14}$C record and suggested two distinct types of the grand minima: shorter
 Maunder-type and longer Sp\"orer-like minima (cf. \cite{stui91}).
Later Voss et al. (1996) defined grand minima in a similar manner
 and provided a list of 29 such events for the last 8000 years.
A similar analysis of excursions of the $^{14}$C production rate has been
 presented by Goslar (2003).
However, because of the lack of adequate physical
 models relating the radicarbon abundance to the solar activity level, such
 studies retained a qualitative element.
The use of high-pass filtered $^{14}$C data is based on the assumption that
 solar activity variations are important only on short times, while
 all the long-term changes in radiocarbon production are attributed solely to
 the slowly changing geomagnetic field.
This method ignores any possible long-term changes in the solar activity
 (e.g., on time scales longer than 500 years for Voss et al. 1996).
There is, however, increasing evidence that solar activity varies on multi-centennial to multi-millennial
 time scales (\cite{mccr04,usos06}).
A recently developed approach, based on physics-based modelling of all
 links relating the measured cosmogenic isotope abundance to the level of
 solar activity, allows for quantitative reconstruction of the solar activity
 level in the past, and thus, for a more realistic definition of
 the periods of grand minima or maxima.

Here we study the statistics of occurrence of grand minima/maxima throughout the
 Holocene and impose additional observational constraints on
 the dynamo models of the Sun and Sun-like stars.

\section{Past solar activity}

\begin{figure}[t]
\resizebox{\hsize}{!}{\includegraphics{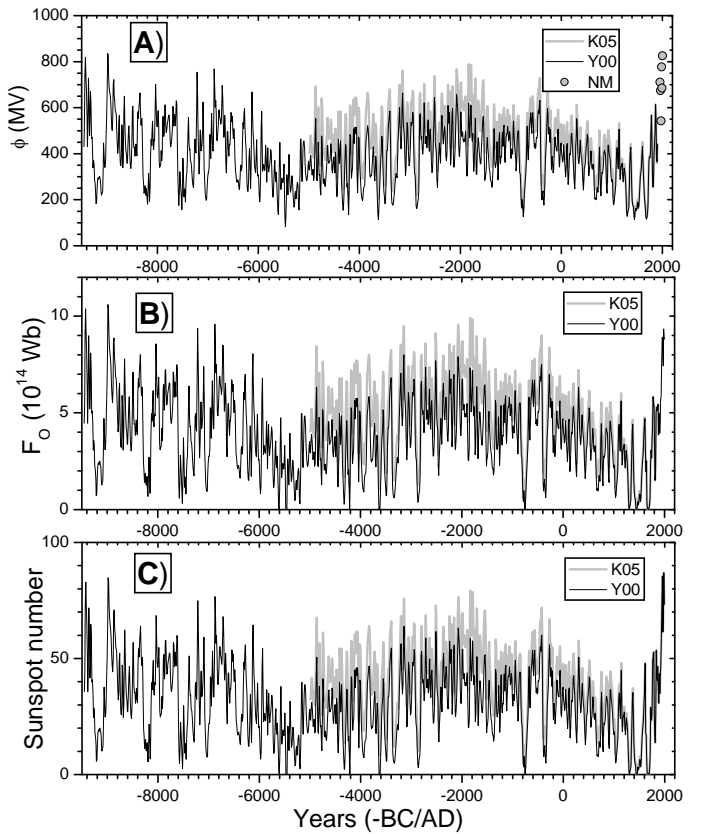}}
     \caption{
     Long-term solar activity reconstruction from $^{14}$C data.
     All data are decadal averages.
     Solid (denoted as 'Y00') and grey ('K05') curves are based on the paleo-geomagnetic
        reconstructions of Yang et al. (2000) and Korte \& Constable (2005), respectively.
     A) The modulation potential $\phi$. Big circles ('NM') denote the $\phi$ values
       obtained from direct cosmic ray measurements since 1951 (\cite{usos05}).
     B) Open flux $F_{\rm o}$. Reconstruction from sunspot numbers (\cite{balm07}) is used after 1610.
     C) Sunspot numbers reconstructed here.
        The Y00 and K05 curves are called SN-L and SN-S series, respectively throughout the paper.
        Observed group sunspot numbers (\cite{hoyt98}) are
        shown after 1610.
      }
     \label{Fig:long}
\end{figure}

Solar activity on multi-millenial time scales has been recently reconstructed
 using a physics-based model from measurements of $^{14}$C in tree rings
 (see full details in \cite{sola04,usos06}).
The validity of the model results for the last centennia has been proven by
 independent data on measurements of $^{44}$Ti in stony meteorites (\cite{usos06b}).
The reconstruction depends on the knowledge of temporal changes of the
 geomagnetic dipole field, which must be estimated independently by paleomagnetic
 methods.
Here we compare two solar activity reconstructions, which are based on alternative
 paleomagnetic models: one which yields an estimate of the virtual aligned dipole moment (VADM)
 since 9500 BC (\cite{yang00}), and the other a recent paleomagnetic
 reconstruction of the true dipole moment since 5000 BC (\cite{kort05}).
We note that the geomagnetic dipole moment obtained by Korte \& Constable (2005) lies systematically lower
 than that of Yang et al. (2000), leading to a systematically higher solar activity reconstruction in the past
 (\cite{usos06}).
While the geomagnetic reconstruction of the VADM by Yang et al. (2000)
 provides an upper bound for the true dipole moment, the more recent work of Korte \& Constable (2005)
 may underestimate it.
Thus we consider both models as they bound a realistic case.
We note that the Yang et al. (2000) data run more than 4000 years longer and
 give a more conservative estimate of the grand maxima.

Different indices of the reconstructed solar activity are shown in Fig.~\ref{Fig:long}.
Most directly related to the $^{14}$C production in the atmosphere is the modulation
 potential $\phi$ (e.g., \cite{cast80,masa99,usos02}) whose variations are shown in panel A.
The modulation potential $\phi$ is a parameter describing the spectrum of galactic
 cosmic rays (see definition and full description of this index in \cite{usos05}).
Using a model of the heliospheric transport of cosmic rays, the modulation potential
 can be nearly linearly related to the open solar magnetic flux $F_{\rm o}$.
The reconstructed long-term variations of the open solar magnetic flux are shown
 in panel B.
Next, using a model of the open magnetic flux formation, one can
 estimate the sunspot numbers from the $F_{\rm o}$ data.
This was done earlier using the open flux model by Solanki et al. (2000, 2002).
However, the model relating sunspots to the open magnetic flux, has been updated
 recently by Krivova, Balmaceda \& Solanki (2007).
Starting from sunspot numbers (SN) they computed open and total magnetic flux as well as
 the total solar irradiance (from SN and total magnetic flux).
Besides the active regions, the influence of ephemeral active regions is included.
They determined the free parameters of the models used by requiring the model output to
 reproduce the best available data sets (of open and total fluxes as well as total solar irradiance)
  with the help of a genetic algorithm.
In particular, the improved total flux data set of Arge et al. (2002) and the total
 irradiance composite of Fr\"ohlich (2006) set tight constraints, which could only be met by revising the relationship
 between SN and total emergent magnetic flux.
This revised relationship is employed here too.
Accordingly, here we use this updated model to convert $F_{\rm o}$ into the
 sunspot number (Eq.~\ref{Eq:est}), which is described in detail in Appendix A.

\begin{figure}[t]
\resizebox{\hsize}{!}{\includegraphics{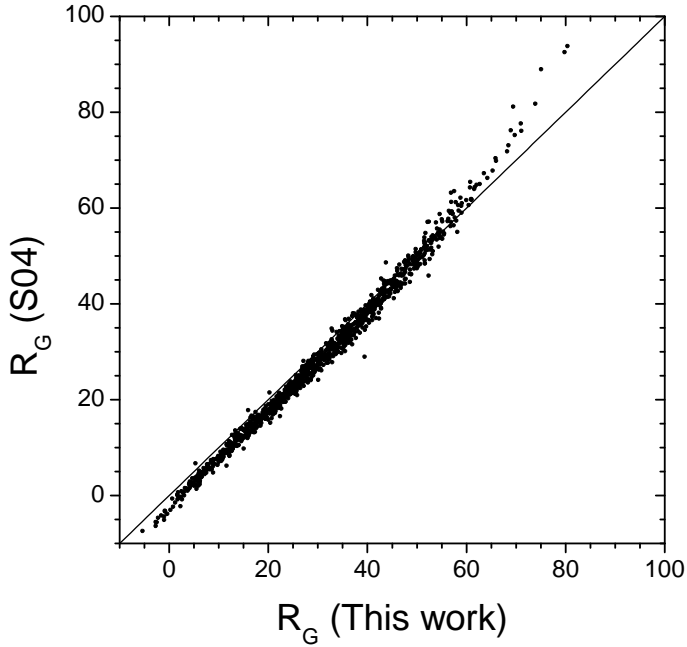}}
     \caption{
     Scatter plot of decadal sunspot numbers for 9500 BC -- 1900 AD published
      by Solanki et al. (2004), $R_{\rm G}$(S04), vs. sunspot numbers obtained in
      this work in a way identical to S04 but using the updated open solar flux model
      of Krivova, Balmaceda \& Solanki (2007).
      }
     \label{Fig:R_R}
\end{figure}

A comparison between the sunspot number series obtained using the new open flux
 model with those published by Solanki et al. (2004) is shown in Fig.~\ref{Fig:R_R}.
The scatter in the relationship is very small (correlation coefficient 0.995),
 but there is a small systematic difference between the two series.
The newly obtained sunspot numbers are slightly higher, with
 the mean difference being about 1.6 and the standard deviation about 2.
At large SN values, however, the opposite is observed, with the new reconstructed SN
 being smaller by values up to 14.
This reduces the largest peaks (grand maxima) somewhat.
This difference is a result of the different relationship between SN
 and total emerging magnetic flux in active regions used by Krivova, Balmaceda \&
 Solanki (2007) than by Solanki et al. (2000, 2002).
We note that the difference is kept within the uncertainty of the
 reconstruction, which is about 8 for the last millenia and up to 15-20
 in the beginning of the Holocene (see Supplementary Material to \cite{sola04}).
The difference between the original sunspot numbers published in Usoskin et al. (2006a)
 based on the paleomagnetic model by Korte \& Constable (2005)
 and those obtained here using the updated open flux model has a mean of 1.0 and
 standard deviation about 2.4.

The 11,000-yr long data sets of the decadal sunspot number
 similar to that by Solanki et al. (2004) but with the updated open flux model
 is shown in Fig.~\ref{Fig:long}C.
It is called the SN-L series throughout the paper.
The shorter series (Fig.~\ref{Fig:long}C), which is similar to
 that by Usoskin et al. (2006), is called SN-S henceforth.
After 1610 AD, the actually observed group sunspot numbers (\cite{hoyt98}) has been
 used instead of the reconstructions.

Before identifying the grand minima and maxima, the decadal resolution data have been smoothed
 with the Gleissberg (1-2-2-2-1) filter, which is regularly applied when studying
 long-term variations of solar activity in order to suppress the noise (e.g., \cite{glei44,soon96,murs98}).
In order to check the effect of the filter we have studied a number of artificial SN series containing a total of 1000 grand
 minima of 60-yr duration (at the level of SN$<$15) each.
A noise with $\sigma=10$ has been added to the series, and the grand minima have been identified
 again as SN$<$15 in both the raw and 1-2-2-2-1 smoothed noised series.
We found that 35\% of grand minima are incorrectly identified (too short, too long or split in two
 short episodes, comparing to the "real" signal) in the raw noisy series.
The filtering reduces the mis-identification to 13$\pm$3\%, i.e. 3-fold.
Thus, the use of the 1-2-2-2-1 Gleissberg filter reduces the effect of noise on the
 grand minima/maxima definition and makes the results more robust.
This smoothing, however, leads to a reduction in the amplitude and a slight underestimate (about 7\%
 according to the above numerical experiment) of the number and duration of short,
 less than 30 years long minima and maxima.

\begin{figure}[t]
\resizebox{\hsize}{!}{\includegraphics{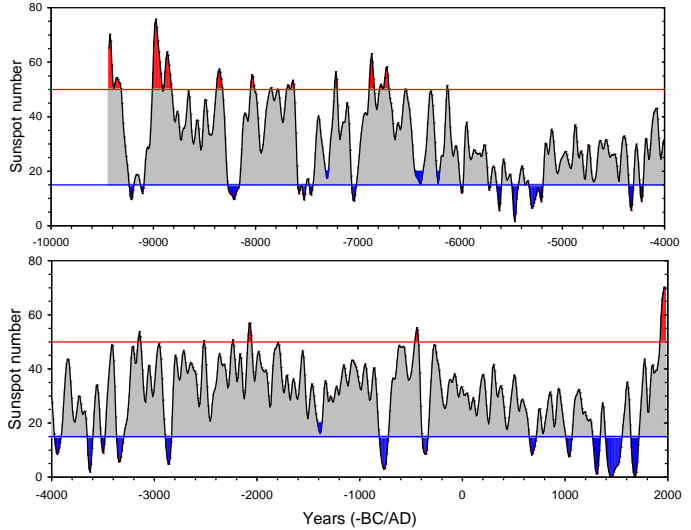}}
     \caption{
     Sunspot activity SN-L throughout the Holocene (see text) smoothed with a 1-2-2-2-1 filter.
     Blue and red areas denote grand minima and maxima, respectively.
     The entire series is spread over two panels for better visibility.
      }
     \label{Fig:SN_S04}
\end{figure}

\begin{figure}[t]
\resizebox{\hsize}{!}{\includegraphics{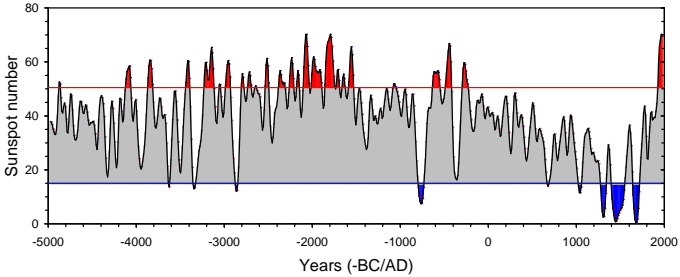}}
     \caption{
     Sunspot activity SN-S (see text) smoothed with a 1-2-2-2-1 filter.
     Blue and red areas denote grand minima and maxima, respectively.
     }
     \label{Fig:SN_U06}
\end{figure}

The filtered SN-L and SN-S series are shown in Figs.~\ref{Fig:SN_S04} and \ref{Fig:SN_U06}, respectively.
We analyze both reconstructed SN data sets in equal details.
%

\section{Definitions}
\subsection{Grand minima}
\label{Sec:mi}
We have defined a grand minimum as a period when the (smoothed) SN level
 is less than 15 during at least two consecutive decades - this corresponds to
 blue-filled areas in Figs.~\ref{Fig:SN_S04} and \ref{Fig:SN_U06}.
However, taking into account the uncertainty of the SN reconstruction and the influence
 of the filtering, we have also considered as minima
 clear dips in the SN whose bottom is between 15 and 20 and the depth
  (with respect to surrounding plateaus/maxima) exceeds 20.
Therefore, such periods as, e.g., ca. 6400 BC are considered as
 grand minima (see Fig.~\ref{Fig:SN_S04}) even though their bottoms are above 15.
On the other hand, the period ca. 4450 BC is not counted as a grand minimum
 because its minimum lies above 15 and its depth is less than 20.
All 27 grand minimum periods thus defined in the SN-L series are
 listed in Table~\ref{Tab:min} together with their approximate duration,
 defined as the period of time when SN was below 15 (20) as discussed above.
Among them there are two periods (ca 9200 BC and 7500 BC) when the dominant
 grand minima are interrupted by a 1-2 decades long upward excursions.
We regard these period as continuous Sp\"orer-like minima.
Together these grand minima have a total duration of 1880 years, so that the Sun spends
 about 17\% of the time in a grand minimum state.

We note that all the grand minima after 3000 BC discussed by Eddy (1977a, 1977b)
 are present in Table~\ref{Tab:min}, which however contains also minima at ca. 1040 BC and
 2860 BC not found by Eddy.
On the other hand, all the grand minima listed in the Table are mentioned
 by Voss et al. (1996), but the latter\footnote{Definition of minima
 by Voss et al. (1996) was based solely on the relative variations of $\Delta^{14}$C.}
 list more minima, e.g., three
 minima between 200 BC and 200 AD which do not appear in our series.
Most of the minima listed here can also bee seen in Fig.~2 of Goslar (2003).
Therefore, we conclude that our definition of grand minima applied to the
 present data set gives results generally in agreement
 with earlier studies but not identical to them.
In particular, it gives more details than the study by Eddy (1977a)
 but discards some small fluctuations mentioned by Voss et al. (1996).
\begin{table}
\caption{Approximate dates (in -BC/AD) of grand minima in the SN-L series (see text).}
\begin{tabular}{cllc}
\hline
No.& center & duration & comment\\
\hline
1 & 1680 & 80 & Maunder \\
2 & 1470 & 160 & Sp\"orer \\
3 & 1305 & 70 & Wolf \\
4 & 1040 & 60 &  a) \\
5 & 685 & 70 & b) \\
6 & -360 & 60 & a,b,c) \\
7 & -765 & 90 & a,b,c) \\
8 & -1390 & 40 & b,e) \\
9 & -2860 & 60 & a,c) \\
10 & -3335 & 70 & a,b,c) \\
11 & -3500 & 40 & a,b,c) \\
12 & -3625 & 50 & a,b) \\
13 & -3940 & 60 & a,c) \\
14 & -4225 & 30 & c) \\
15 & -4325 & 50 & a,c) \\
16 & -5260 & 140 & a,b) \\
17 & -5460 & 60 & c) \\
18 & -5620 & 40 & - \\
19 & -5710 & 20 & c) \\
20 & -5985 & 30 & a,c) \\
21 & -6215 & 30 & c,d) \\
22 & -6400 & 80 & a,c,d) \\
23 & -7035 & 50 & a,c) \\
24 & -7305 & 30 & c) \\
25 & -7515 & 150 & a,c) \\
26 & -8215 & 110 & - \\
27 & -9165 & 150 & - \\
\hline
\end{tabular}
\\
a) Discussed in Stuiver (1980) and Stuiver \& Braziunas (1989).\\
b) Discussed in Eddy (1977a, 1977b).\\
c) Shown in Goslar (2003).\\
d) Exact duration is uncertain.\\
e) Does not appear in the SN-S series.
 \label{Tab:min}
\end{table}

The grand minima listed in Table~\ref{Tab:min} dating after 5000 BC are identical for both
 SN-L and SN-S series (except for one minimum ca. 1385 BC in the SN-S series which does not
 match the formal definition).
These grand minima will be used for the further analysis.

\subsection{Grand maxima}

Similar to Solanki et al. (2004), we define as a grand maximum of solar activity a period when SN exceeds 50
 during at least two consecutive decades (see red filled areas in Figs.~\ref{Fig:SN_S04}
 and \ref{Fig:SN_U06}).
If two consecutive maxima are separated by less than 30 years they are considered as a single
 maximum (e.g., ca. 9000 BC in the SN-L series), i.e. they are treated in a way similar to grand minima.
We have identified 19 grand maxima (of a total duration of 1030 years, corresponding to about 9\% of the time)
 in the SN-L series
 since 9500 BC, including also the modern maximum.
These are listed in Table~\ref{Tab:max}.
Four out of six grand maxima found here after 3000 BC coincide with those pointed out
 by Eddy (1977a,1977b).
In the SN-S series, 23 grand maxima (of a total duration of 1560 years, corresponding to
 about 22\% of the time) are identified since
 5000 BC, as listed in Table~\ref{Tab:max_K}.
All maxima identified in the SN-L series are present also in the SN-S series, but
 the latter yields more maxima satisfying the same definition before
 1500 BC (after ca. 1500 BC the maxima are nearly identical).
This indicates that the identification of maxima is less robust than for grand minima,
 and is more dependent on the definitions and model assumptions.
\begin{table}
\caption{Approximate dates (in -BC/AD) of grand maxima in the SN-L series.}
\begin{tabular}{cllc}
\hline
No.& center & duration & comment\\
\hline
1   &   1960    &   80  &   modern, b)  \\
2   &   -445    &   40  &   -   \\
3   &   -1790   &   20  &   a)  \\
4   &   -2070   &   40  &   -   \\
5   &   -2240   &   20  &   a)  \\
6   &   -2520   &   20  &   a)  \\
7   &   -3145   &   30  &   -   \\
8  &   -6125   &   20  &   -   \\
9  &   -6530   &   20  &   -   \\
10  &   -6740   &   100  &   -   \\
11  &   -6865   &   50  &   -   \\
12  &   -7215   &   30  &   -   \\
13  &   -7660   &   80  &   -   \\
14  &   -7780   &   20  &   -   \\
15  &   -7850   &   20  &   -   \\
16  &   -8030   &   50  &   -   \\
17  &   -8350   &   70  &   -   \\
18  &   -8915   &   190  &  -   \\
19  &   -9375   &   130 &   -   \\
\hline
\end{tabular}
\\
a) Discussed in Eddy (1977a, 1977b)).\\
b) Center and duration of the modern maximum are preliminary since it is
 still ongoing.\\
 \label{Tab:max}
\end{table}

\begin{table}
\caption{Approximate dates (in -BC/AD) of grand maxima in the SN-S series.}
\begin{tabular}{cllcc}
\hline
No.& center & duration & comment\\
\hline
1   &   1960    &   60  &   a,b)   &   \\
2   &   -265    &   70  &   -   &   \\
3   &   -455    &   70  &   a)   &   \\
4   &   -595    &   90  &   -   &   \\
5   &   -1065   &   50  &   -   &   \\
6   &   -1560   &   60  &   -   &   \\
7   &   -1640   &   40  &   -   &   \\
8   &   -1775   &   170 &   a)   &   \\
9   &   -1995   &   210 &   a)   &   \\
10  &   -2165   &   30  &   -   &   \\
11  &   -2235   &   50  &   a)   &   \\
12  &   -2350   &   80  &   -   &   \\
13  &   -2515   &   50  &   a)   &   \\
14  &   -2645   &   30  &   -   &   \\
15  &   -2715   &   50  &   -   &   \\
16  &   -2790   &   40  &   -   &   \\
17  &   -2960   &   60  &   -   &   \\
18  &   -3030   &   40  &   -   &   \\
19  &   -3170   &   120 &   a)   &   \\
20  &   -3415   &   50  &   a)   &   \\
21  &   -3840   &   60  &   -   &   \\
22  &   -4090   &   60  &   -   &   \\
23  &   -4870   &   20  &   -   &   \\
\hline
\end{tabular}
\\
a) Exists also in the SN-L series (Table~\ref{Tab:max}).\\
b) Center and duration of the modern maximum are preliminary since it is
 still ongoing.\\
 \label{Tab:max_K}
\end{table}

\subsection{Waiting time distribution}
\label{Sect:power}

The interval between two consequent events is called the waiting time.
The statistical distribution of waiting times (WTD -- waiting time distribution)
 reflects the nature of a process which produces the studied events.
For instance, an exponential WTD is a clear indicator of a purely random,
 "memoryless" process (e.g., Poisson process), when the behaviour of a system
 does not depend on its preceding evolution on both short or long time-scales.
Any significant deviation of the WTD from an exponential law
 implies that the probability of an event to occur is not time-independent but
 is related to the previous history of the system.
We note that the occurrence of events generally is random also for a non-exponential
 distribution, but the probability is not uniform in time.
This can be interpreted in different ways:
 self-organized criticality (e.g., \cite{carv00,free00}), time-dependent
 Poisson process (e.g., \cite{whea03}), some memory in the driving process
 (e.g., \cite{lepr01,mega03}).
The most typical non-exponential WTD is a power law which is, e.g., a necessary but not sufficient
 indication of self-organized criticality (\cite{carv00}).
A power law implies higher tails of the distribution, i.e.
 higher probability (relative to the exponential function) of occurrence of both long and
 short intervals between the events.
A power law distribution of the waiting time has been obtained for
 many solar and terrestrial indices on different time scales from minutes
 to 10$^5$ years:
 e.g., intervals between major earthquakes (\cite{bak02,mega03});
 intervals between successive solar flares (\cite{pear93,boff99,moon01});
 waiting time between successive coronal mass ejections (\cite{whea03,berh06});
 intervals between bursts in the solar wind (\cite{free00});
 repetition time of geomagnetic disturbances (\cite{papa06});
 intervals between the geomagnetic field reversals (\cite{pont06}), etc.
Note that many of these processes, which depict different degrees of
 self-organization, are related to energy accumulation and
 release.

Here we study the WTD of the occurrence of grand minima and maxima
 of solar activity in order to understand the nature of its long-term evolution.
The waiting time is defined as the
 length $x$ of an interval between centers of consequent events.

First we studied the {\it differential distribution} which is
 defined as
\begin{equation}
y(x)={N\{x_1,x_2\}\over x_2-x_1},
\label{Eq:dif}
\end{equation}
 where $N\{x_1,x_2\}$ is the number of events with the
 waiting time $x_1\leq x<x_2$.
Statistical errors of the differential distribution are estimated from
 the Poisson statistics as $\sqrt{N}/(x_2-x_1)$.

Since the statistics are poor (19--27 events) and differential WTD histograms are rough,
 we have studied also the normalized {\it cumulative distribution} defined as
\begin{equation}
 Y(x)={N\{x,\infty\}\over N\{0,\infty\}},
\label{Eq:int}
\end{equation}
which corresponds to the probability of finding a waiting time exceeding $x$.
In this case the statistical errors cannot be determined since the points of the
 cumulative distribution are not independent.

The {\it exponential} WTD model is defined as
\begin{equation}
y(x)\propto \exp{\left({-x\over\tau}\right)};\hskip 0.3cm
Y(x)\propto \exp{\left({-x\over{\rm T}}\right)};\hskip 0.3cm
\tau={\rm T}.
\label{Eq:exp}
\end{equation}
The {\it power law} WTD model is defined as
\begin{equation}
y(x)\propto x^{-\gamma};\hskip 0.4cm
Y(x)\propto x^{-\Gamma};\hskip 0.4cm
\gamma=\Gamma+1.
\label{Eq:PL}
\end{equation}
The power law serves mainly to emphasize deviations from a purely exponential
 distribution, since the poor statistics hardly allow us to distinguish between a power law
 and other more-stretched-than-exponential distributions, e.g., log-Poisson.

We note that short intervals (shorter than a century) cannot be reliably
 defined because of noise and filtering.
Statistics of very long intervals is not reliable
 either because of the limited length of the analyzed series.
Therefore, when fitting the data we will ignore the shortest and longest intervals,
 i.e. first and last points of the cumulative distribution (the number of bins of the differential
  distribution is left unchanged).

\section{Analysis and results}
\subsection{Sunspot number distribution}

\begin{figure}[t]
\resizebox{\hsize}{!}{\includegraphics{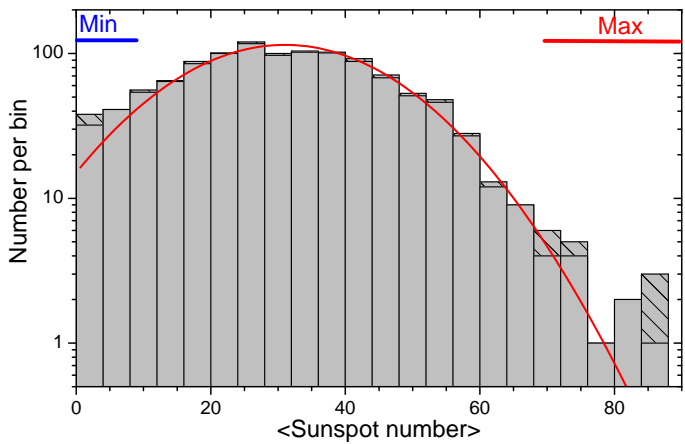}}
     \caption{
     Histogram of the sunspot number SN series reconstructed here
      for 9,500 BC -- 2000 AD.
     Hatched areas correspond to directly observed sunspots after 1610.
     The curve represents the best fit normal distribution.
     }
     \label{Fig:SN_hist}
\end{figure}

First we have constructed histograms of the sunspot numbers.
The histogram for the SN-L series is shown in Fig.~\ref{Fig:SN_hist}.
While being close to a normal distribution
 (mean=31, $\sigma=30$), there is an apparent excess both at
 very low sunspot numbers, corresponding to the grand minima, and
 at very high values, corresponding
 to grand maxima.
The overall distribution is consistent with the direct observational
 record after 1610, suggesting that the latter is a representative
 sample for the sunspot activity statistics, including grand minimum
 and maximum \footnote{Note that in Fig.~\ref{Fig:SN_hist} intermediate SN values are
 seemingly underepresented in modern times, but this is due to the
 logarithmic scale.}.
This distribution with these excesses suggests that grand minima and maxima are special
 states of the solar dynamo that cannot be explained by random
 fluctuations or noise in the data (see also forthcoming sections).

\begin{table*}
\caption{Fitting of power law and exponential models to distributions of the grand minima
 and maxima occurrence:
 For the differential distribution the value of $\chi^2$ is shown together
  with the corresponding confidence level (in parentheses) for 4 degrees of freedom. }
\begin{tabular}{p{2cm}|p{6cm}|p{6cm}}
\hline
 & \hskip 1.5cm Differential distribution& \hskip 1.5cm Cumulative distribution\\
\end{tabular}
\begin{tabular}{p{2cm}|p{2.8cm}|p{2.78cm}|p{2.8cm}|p{2.8cm}}
 Series & Power law & Exponential & Power law & Exponential\\
 \hline
 {\bf A)} SN-L & $\gamma=1.61\pm 0.1$ & $\tau=330\pm 50$ yr & $\Gamma=0.95\pm 0.02$ & $T=435\pm 15$ yr\\
 min WTD &$\chi^2=0.35$ (0.99) & $\chi^2=2.2$ (0.7) & & \\
 \hline
 {\bf B)} SN-L & $\gamma=1.36\pm 0.1$ & $\tau=430\pm 30$ yr & $\Gamma=0.77\pm 0.05$ & $T=355\pm 20$ yr\\
 max WTD &$\chi^2=0.26$ (0.992) & $\chi^2=1.8$ (0.79) & & \\
 \hline
 {\bf C)} SN-S & $\gamma=1.82\pm 0.06$ & $\tau=250\pm 40$ yr & $\Gamma=0.95\pm 0.04$ & $T=290\pm 25$ yr\\
 max WTD &$\chi^2=0.22$ (0.994) & $\chi^2=6.5$ (0.16) & & \\
 \hline
 {\bf D)} SN-L & $\gamma=1.25\pm 0.18$ & $\tau=51\pm 5$ yr & $\Gamma=1.22\pm 0.12$ & $T=55\pm 2$ yr\\
 max duration &$\chi^2=1.06$ (0.9) & $\chi^2=0.58$ (0.97) & & \\
 \hline
 {\bf E)} SN-S & $\gamma=1.5\pm 0.6$ & $\tau=50\pm 10$ yr & $\Gamma=1.44\pm 0.14$ & $T=59\pm 6$ yr\\
 max duration &$\chi^2=7.0$ (0.14) & $\chi^2=3.3$ (0.51) & & \\
 \hline
\end{tabular}
\label{Tab:RES}
\end{table*}

\subsection{Grand minima}

\subsubsection{Waiting time distribution}

\begin{figure}[t]
\resizebox{\hsize}{!}{\includegraphics{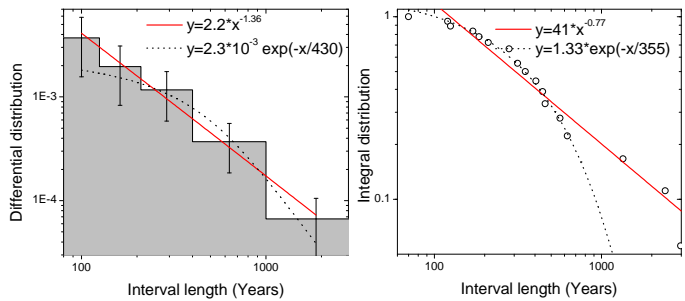}}
     \caption{
     Differential (left panel) and cumulative (right panel)
      distribution of the waiting time between subsequent grand minima.
     The histogram (left) and circles (right) represent the observed distribution, while solid and dotted
      lines depict best fit power law and exponential approximations, respectively.
     }
     \label{Fig:dist}
\end{figure}

The distribution of the waiting time between grand minima is shown in
 the left-hand panel of Fig.~\ref{Fig:dist} together
 with the best fit approximations.
Parameters of the best-fit approximations are shown in Table~\ref{Tab:RES} (row A).
The best fit exponential model (Eq.~\ref{Eq:exp}) gives
 $\tau=330\pm 50$ years, which roughly corresponds to the mean frequency of
 grand minima occurrence.
The exponential model agrees only relatively poorly with the observed WTD.
The best fit power law model (Eq.~\ref{Eq:PL}) agrees reasonably with the observed WTD.

The cumulative WTD is shown in the right-hand panel of
 Fig.~\ref{Fig:dist} together with the best fit approximations (Table~\ref{Tab:RES}, row A,
  columns 4 and 5).
The power law model agrees well with the bulk of the data
 except for the very far tail ($x>$1000 years).
However, this tail contains only two events and is not representative.
(The $\chi^2$-test cannot be applied to the cumulative distribution since the points
 are not independent.)
The exponential model poorly reproduces the WTD.
As an additional test we compare the parameters of the models describing differential
 and cumulative distributions, viz. T$\approx\tau$ and $\gamma\approx\Gamma+1$.
Both models pass this test.

We conclude that a power law model better describes the observed WTD for
 grand minima, although an exponential decay cannot be completely ruled out.
This is valid also for the SN-S series whose grand minima (except of one ca. 1385 BC) coincide
 with those in the SN-L series after 5000 BC.

\subsubsection{Duration of grand minima}

\begin{figure}[t]
\resizebox{\hsize}{!}{\includegraphics{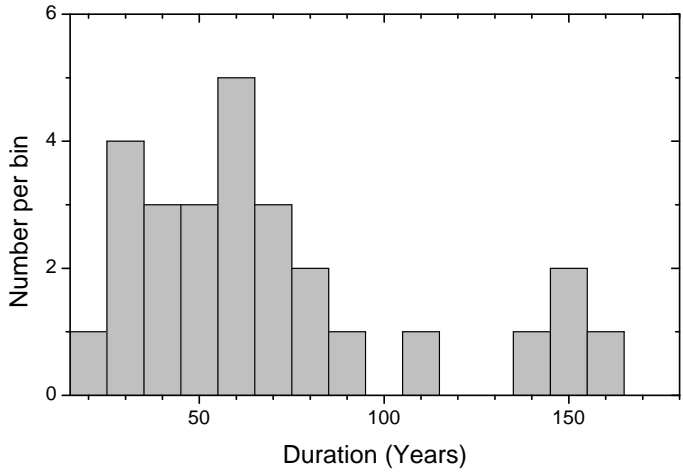}}
     \caption{
     Histogram of the duration of grand minima.
     }
     \label{Fig:dur}
\end{figure}

A histogram of the duration of grand minima (Table~\ref{Tab:min}) is shown in Fig.~\ref{Fig:dur}.
The mean duration is 70 year but the distribution is not uniform.
The minima tend to be either of a short duration, between 30 and 90 years
 similar to the Maunder minimum, or rather long, longer than 110 years
 similar to the Sp\"orer minimum.
This agrees with the earlier conclusion on two different types of grand minima (\cite{stui89,gosl03}).
This suggests that a grand minimum is a special state of the dynamo
 whose duration is not random but is defined by some intrinsic process.
Note, however, that only 3 of the 5 Sp\"orer-like minima are clear long grand minima while
 the other 2 are composed of multiple sub-minima (\# 25 and 27 in Table~\ref{Tab:min} -- see Sect.~\ref{Sec:mi}).

\subsection{Grand maxima}

\subsubsection{Waiting time distribution}

\begin{figure}[t]
\resizebox{\hsize}{!}{\includegraphics{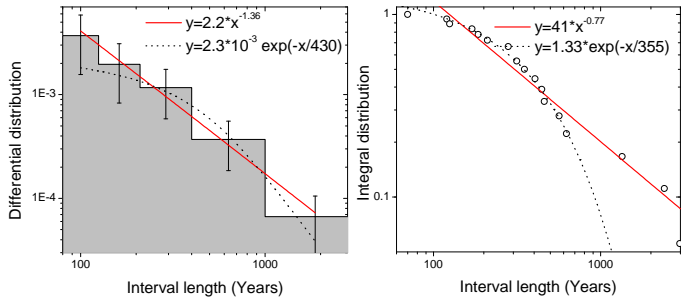}}
     \caption{
     Differential (left panel) and cumulative (right panel) distributions
     of the waiting time between grand maxima according to the SN-L series.
     }
     \label{Fig:ma_int_50}
\end{figure}

The distribution of the waiting time intervals between
 subsequent maxima, listed in Table~\ref{Tab:max}, is shown
 in Fig.~\ref{Fig:ma_int_50}, with the parameters of best-fit approximations
 shown in Table~\ref{Tab:RES}, row B.
The differential distribution (left panel) can be well
 fitted by the power law model.
An exponential model gives a formally insignificant
 fit to the distribution.

The cumulative distribution is shown in the right panel and is
 also close to a power law (see Table~\ref{Tab:RES}, rowB).
The exponential model fits
 short-to-long intervals even better, but cannot reproduce the far tail,
 with three intervals exceeding 1000 years.
Indices for the differential and cumulative models are
 barely consistent with each other (T$\approx\tau$ and $\gamma\approx\Gamma+1$).
Accordingly, for the SN-L series we cannot give a clear preference to either model.

\begin{figure}[t]
\resizebox{\hsize}{!}{\includegraphics{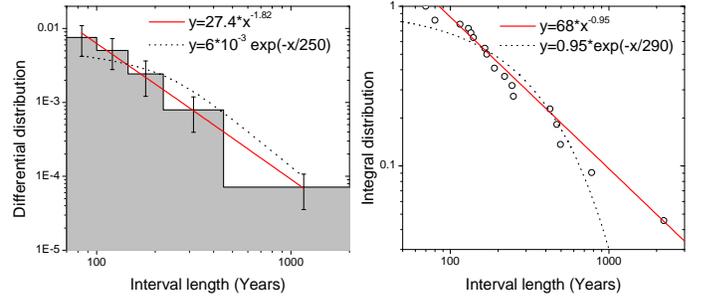}}
     \caption{
     Differential (left panel) and cumulative (right panel) distributions
     of the waiting time between grand maxima according to the SN-S series.
     }
     \label{Fig:max_int_K}
\end{figure}

The statistics of WTD for the grand maxima using the SN-S series is shown in Fig.~\ref{Fig:max_int_K},
 with the best-fit parameters listed in Table~\ref{Tab:RES}, row C.
The power law model satisfactory fits the differential WTD,
 while the exponential law displays only a poor correspondence to it.
The cumulative WTD (right  panel) is nicely fitted by a power law but poorly by an exponential.
Both models pass the additional test ($\gamma\approx\Gamma+1$ vs. $T\approx\tau$).

Therefore we conclude that, although the exponential model cannot be
 totally excluded, the power law model is more preferable in describing the WTD of
 grand maxima.

\subsubsection{Duration of grand maxima}

\begin{figure}[t]
\resizebox{\hsize}{!}{\includegraphics{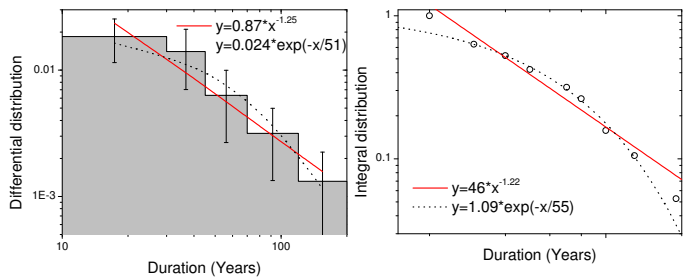}}
     \caption{
     Histogram of the duration of grand maxima.
     }
     \label{Fig:ma_du_50}
\end{figure}

The distribution of the lengths of maxima in the SN-L series is shown in Fig.~\ref{Fig:ma_du_50},
 with best-fit parameters listed in Table~\ref{Tab:RES}, row D.
The differential distribution (left panel) is reasonably fitted by an exponential law
 but is poorly described by a power law.
Although both models are seemingly good in fitting the observed cumulative WTD (right panel of
 Fig.~\ref{Fig:ma_du_50}), the additional test excludes the power law model, since
 T$\approx\tau$ but $\gamma\neq\Gamma+1$.
Therefore, we conclude that the distribution of the lengths of grand maximum
 episodes is close to exponential, as noticed by Solanki et al. (2004).

The differential distribution of the duration of maxima for the SN-S series is fitted
 by none of the two models (see Table~\ref{Tab:RES}, row E).
The best-fit parameters for the cumulative distribution also favor the exponential
 distribution ($\tau\approx T$) over the power law ($\Gamma\neq\gamma+1$).

\subsection{Quasi-periodicities}

We have also studied possible quasi-periodicities in the rate of grand minima/maxima
 occurrence.
We have found that the occurrence of grand minima depicts a weak (marginally
 significant) quasi-periodicity of 2000-2400 years, which is a well-known period
 in $^{14}$C data (e.g., \cite{damo91,vasi02}).
No other periodicities are observed in the occurrence rate of grand minima.
We have found no periodic feature in the occurrence of grand maxima in the
 SN-L series, while a marginal hint for a periodicity of about 1200 years and its harmonics
 (about 600 and 400 years - cf. \cite{usos04}) is found in SN-S data.
This indicates that the 2400-year periodicity is related likely to the clustering of
 grand minima rather than to a long-term "modulation" of solar activity.
Therefore, we conclude that the occurrence of grand minima and maxima is not
 a result of long-term cyclic variability but is defined by stochastic/chaotic
 processes as discussed in Sect.~\ref{Sec:dis}.

\section{Summary of the results}
\label{Sect:Sum}

We have studied the statistics of occurrence of grand minima and maxima
 over the last 7--11 millennia.
The main results can be summarized as follows.
\begin{enumerate}

\item
We have presented lists of grand minima (Table~\ref{Tab:min}) and maxima
 (Tables~\ref{Tab:max} and \ref{Tab:max_K}), using updated physics-based reconstruction
 of solar activity from $^{14}$C data measured by the INTCAL collaboration (\cite{stui98}).
The identification of grand minima is found to be more robust to the exact correction of the
 geomagnetic field than the grand maxima.

\item
\label{It:per}
The occurrence of grand minima and maxima does not depict a dominant periodic behaviour.
Only a weak tendency exists for grand minima to cluster with a quasi-period
  of about 2400 years, and no clear periodicities are observed in the occurrence of grand maxima.

\item
The waiting time between grand minima depicts a distribution which deviates significantly
 from the exponential distribution \footnote{We confronted the data with a power law
 WTD, but can hardly distinguish it from other non-exponential distributions.
 The most important point here is the deviation from a purely exponential WTD.},
 although the latter cannot be completely ruled out
 because of poor statistics.

\item
The distribution of the duration of grand minima is bimodal, with a dominance of short
 (30-90 yr) Maunder-like minima and a smaller number of long (longer than 110 yr)
 Sp\"orer-like minima.

\item
The distribution of the waiting time between grand maxima also deviates from an exponential
 distribution (especially for the last 7000 years in the SN-S data series), but the latter
 cannot be completely ruled out.

\item
Lengths of grand maxima correspond to an exponential distribution.

\end{enumerate}

We have tested that the obtained results are robust with respect
 to the uncertainties of the reconstruction.
The results remain qualitatively the same when varying the parameters
 of Eq.~(\ref{Eq:est}) or when using the sunspot data obtained earlier (\cite{sola04,usos06})
 based on the old open flux model.

\section{Discussion and Conclusions}
\label{Sec:dis}

Using the above results we can formulate additional constraints on a dynamo model
 aiming to describe the long-term evolution of solar magnetic activity.

\begin{enumerate}
\item
The Sun spends around 3/4 of the time at moderate magnetic activity levels (averaged over
 10 years).
The remainder of the time is spent in the state of a grand minimum (about 17\%) or a
 grand maximum (9\% or 22\% for the SN-L or SN-S series, respectively).
The solar activity during modern times corresponds to the grand maximum state.

\item
The occurrence of grand minima/maxima is not a result of long-term cyclic
 variations but is defined by stochastic/chaotic processes.
This casts significant doubts on attempts of a long-term prediction of solar activity
   using multi-periodic analyses.

\item
The observed waiting time distribution of the occurrence of both grand minima and
 grand maxima displays a deviation from an exponential distribution.
A relative excess of short and long waiting times indicates that the occurrence
 of these events is not a time independent "memoryless" Poisson-like process,
 but tends to either cluster events together or produce long event-free periods.
Similar waiting time distributions are typical for many processes with, e.g.
 self-organized criticality or processes related to accumulation and release of energy
 (see Sect.~\ref{Sect:power}).

\item
We distinguish between grand minima of two different types: short minima of Maunder
 type and long minima of Sp\"orer type (cf., \cite{stui89}).
This suggests that a grand minimum is a special state of the dynamo.
Once falling into the grand minimum as a result of a stochastic/chaotic but non-Poisson process,
 the dynamo is "trapped" in this state and its behaviour is driven by
 deterministic intrinsic features.

\item
The duration of grand maxima follows an exponential distribution, in accord with
 the earlier finding of Solanki et al. (2004).
This indicates that leaving a grand maximum is a random process, in contrast to the
 grand minimum case.

\end{enumerate}

In conclusion, we have presented an analysis of the occurrence of grand minima
 and maxima of solar activity on time scales up to 11,000 years.
The results put important observational constraints upon the long-term behaviour of the solar dynamo.
In view of the solar paradigm for the magnetic activity of cool stars, we expect these results
 to be applicable also to stellar dynamo models.
We note, however, that the current results depend on the reliability of the reconstruction
 of the sunspot numbers, which in turn depends on the reliability of the employed geomagnetic
 field and other factors.
This mainly affects the definition of grand maxima, while the statistics of grand minima
 occurrence remain fairly robust against these uncertainties.

\begin{acknowledgements}
Natalie Krivova and Laura Balmaceda are thanked for useful discussions and for
 input that provided the basis for revising
 the sunspot number reconstruction described in the appendix.
We are grateful to Monika Korte, Vincent Courtillot, Gauthier Hulot and Arnaud Chulliat for
 useful discussion on the paleomagnetic data.
GAK gratefully acknowledges supports from the Academy of Finland and the Finnish Academy of Science and
  Letters Vilho, Yrj\"o and Kalle V\"ais\"al\"a Foundation.
\end{acknowledgements}

\begin{appendix}
\section{Conversion between the solar open magnetic flux and sunspot numbers}

Here we invert the updated model relating the sunspot number $R$ to the open magnetic flux $F_o$
 (\cite{balm07} - referred henceforth as KBS07) to reconstruct
 the decadal sunspot numbers from the open flux (cf. \cite{usos04}) as follows.
From Eq. (3) of KBS07 one can obtain (henceforth $\langle ... \rangle$ denotes 10-year averaging):
\begin{equation}
\Big\langle {{{\rm d}F_{o}}\over {{\rm d} t}}\Big\rangle = \langle S\rangle - {\langle F_{o}\rangle\over {\tau_{o}}},
\end{equation}
where
\begin{equation}
\langle S\rangle={\langle F_{act}\rangle\over {\tau_{ta}}}+{\langle F_{eph}\rangle\over {\tau_{te}}}.
\label{Eq:S}
\end{equation}
From Eq. (1) of KBS07 it follows
\begin{equation}
{{{\rm d}F_{act}}\over {{\rm d} t}} = \varepsilon_{act}(t) - {{ F_{act}}\over {{\tau_1}}},
\end{equation}
where
\begin{equation}
{1\over{\tau_1}}={1\over{\tau_{act}}}+{1\over{\tau_{ta}}}.
\end{equation}
Hereafter we adopt the parameter values from Table 1 (line 1) and Sect. 2.1 of KBS07
 and express magnetic flux in units of $10^{14}$ Wb/month and time in months.
Eq. (5) of KBS07 takes the form
\begin{equation}
 \varepsilon_{act}(t) \approx 0.128 R(t),
\label{eq:eps}
\end{equation}
Since $\tau_1\approx 3$ months is much shorter than the cycle length, one can assume that
$F_{act}(t)\approx\tau_1\cdot \varepsilon_{act}(t),$
and after 10-year averaging we obtain:
\begin{equation}
\langle F_{act }\rangle \approx {0.38} \langle R \rangle .
\label{Eq:Fa}
\end{equation}
Similarly, since $\tau_{eph}\approx 0.019$ months is very small, one expects
\begin{equation}
F_{eph}(t)\approx\tau_{eph} \cdot\varepsilon_{eph}(t),
\label{eq:epa}
\end{equation}
where
\begin{equation}
 \varepsilon_{eph}(t)= 117\cdot \varepsilon_{act,max,i}\cdot \sin^2(t').
\end{equation}
Here $t'$ runs over the length of ephemeral region cycle, which is longer than
 a sunspot cycle.
With the actual $R_{\rm G}$ data since 1700 (\cite{hoyt98}) we have tested that the amplitude of
 a solar cycle is related to the 10-yr averaged sunspot value as
\begin{equation}
R_{max}=(2.2\pm 0.4)\langle R\rangle .
\end{equation}
Therefore, from (\ref{eq:eps})
\begin{equation}
\varepsilon_{act,max.i} ={0.128} R_{max,i}\approx (0.28\pm0.05) \langle R \rangle .
\end{equation}
Since $\varepsilon_{eph}$ displays long cycles (of the mean duration of about 18 years)
 that partially overlap, the mean value of $\langle \sin^2(t') \rangle$ is
 numerically found to be 0.74$\pm$0.02, and Eq.~(\ref{eq:epa}) becomes
\begin{equation}
\langle F_{eph}\rangle\approx(0.46\pm0.08) \langle R \rangle .
\label{Eq:Fe}
\end{equation}
From the above consideration one obtains that
$ \langle S\rangle\approx (0.0028\pm 0.0001) \langle R \rangle$
and
\begin{equation}
\langle R\rangle\approx(8.5\pm 0.3)\cdot\langle F_{o}\rangle+(357\pm 13)\cdot\Big\langle {{{\rm d}F_{o}}\over {{\rm d} t}}\Big\rangle .
\label{Eq:R}
\end{equation}
In order to evaluate the decadal mean derivative we substitute the derivative by a slope.
\begin{equation}
\Big\langle {{{\rm d}F(t)}\over {{\rm d} t}}\Big\rangle \approx {\Delta \langle{F}\rangle\over\Delta t} .
\label{Eq:ass}
\end{equation}
However, from decadal data we cannot evaluate the value of $\Delta\langle F\rangle$ over a
 calendar decade:
\begin{equation}
\Delta \langle{F}\rangle_i=\langle{F(t_i+10{\rm yr})}\rangle-\langle{F(t_i)}\rangle,
\end{equation}
where $t_i$ denotes the start year of a decade.
Instead we have to use the value of
\begin{equation}
\langle{F}\rangle_{i+1}-\langle{F}\rangle_i=\langle{F(t_{i+1}+5{\rm yr})}\rangle-\langle{F(t_{i}+5{\rm yr})}\rangle,
\end{equation}
 which is displaced in time with respect to the exact definition of $\Delta \langle{F}\rangle$.
From the open flux computed by KBS07 we have evaluated that
\begin{equation}
\Big\langle{{\rm d}F_{o}\over{\rm d}t}\Big\rangle=(0.73\pm 0.06)
\cdot {\langle F_{o}\rangle_{i+1}-\langle F_{o}\rangle_i\over 120{\rm months}}.
\label{Eq:dF}
\end{equation}
Entering Eq.~(\ref{Eq:dF}) into Eq.~(\ref{Eq:R}), one gets
\begin{equation}
\langle R\rangle_i\approx(8.5\pm 0.3)\cdot\langle F_{o}\rangle_i+(2.2\pm 0.2)\cdot\Big( \langle F_{o}\rangle_{i+1}-\langle F_{o}\rangle_i\Big).
\label{Eq:est}
\end{equation}

\begin{figure}[t]
\resizebox{\hsize}{!}{\includegraphics{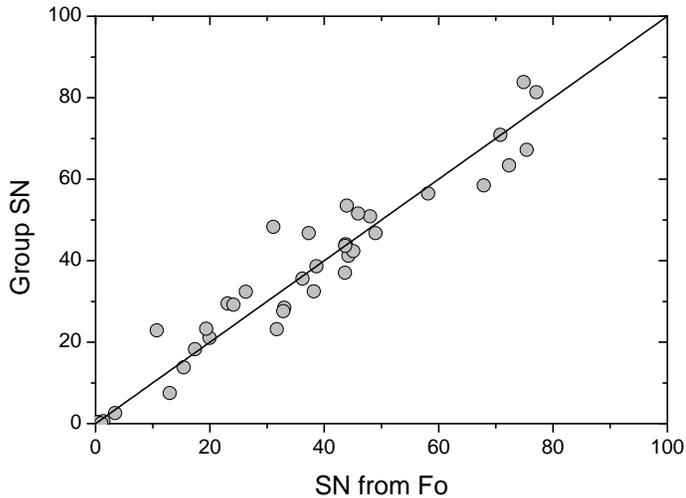}}
     \caption{
     Relation between the decadal group sunspot numbers (\cite{hoyt98})
         and sunspot numbers computed using relation (\ref{Eq:est}) from the open magnetic flux
         (KBS07) for 1610--2000.
       The diagonal, representing the expectation value, is shown by the solid line.
       }
     \label{Fig:F_v_SN}
\end{figure}

We have tested the relation between actual 10-year averaged GSN and SN computed using Eq.~(\ref{Eq:est})
 from the open flux (KBS07) for the period 1611--2000.
The scatter plot shown in Fig.~\ref{Fig:F_v_SN} displays a good correspondence between GSN
 and SN obtained from Eq.~(\ref{Eq:est}).
The cross correlation is $r=0.96$, and the difference displays a nearly Gaussian distribution
 with an offset of -0.4 and $\sigma=5.4$.

Thus, we conclude that the conversion between $F_o$ and SN decadal data can be done
 including the influence of ephemeral regions
 in a straightforward manner via Eq.~(\ref{Eq:est}) with an uncertainty of a few units in sunspot numbers.

\end{appendix}


\begin{thebibliography}{}

\bibitem[Arge et al. 2002]{arge02}
Arge, C.N., Hildner, E., Pizzo, V.J. \& Harvey, J.W., 2002, J. Geophys. Res., 107(A10), 1319.

\bibitem[Bak et al. 2002]{bak02}
Bak, P., Christensen, K., Danon, L. \& Scanlon, T., 2002,
 Phys. Rev. Lett., 88, 178501

\bibitem[Damon \& Sonett 1991]{damo91}
Damon, P.E. \& Sonett, C.P. 1991, in {\it Sun in Time}, ed. C.P. Sonett, M.S. Giampapa \& M.S. Matthews (Tucson, Univ. Arizona Press), 360.

\bibitem[Berhondo et al. 2006]{berh06}
Berhondo A.L.M., Taboada, R.E.R. \& Larralde, L.A. 2006, Astrophys. Space Sci.,
 302, 213

\bibitem[Boffetta et al. 1999]{boff99}
Boffetta, G., Carbone, V., Giuliani, P., Vetri, P. \& Vulpiani, A. 1999,
 Phys. Rev. Lett., 83, 4662

\bibitem[Charbonneau 2001]{char01}
Charbonneau, P. 2001, Solar Phys., 199, 385

\bibitem[Charbonneau et al. 2004]{char04}
Charbonneau, P., Blais-Laurier, G. \& St-Jean, C. 2004, ApJ, 616, L183.

\bibitem[de Carvalho \& Prado 2000]{carv00}
de Carvalho, J.X. \& Prado C.P.C 2000,
 Phys. Rev. Lett., 84, 4006

\bibitem[Castagnoli \& Lal 1980]{cast80}
Castagnoli, G. \& Lal, D. 1980, Radiocarbon, 22(2), 133.

\bibitem[Choudhuri 1992]{chou92}
Choudhuri, A.R. 1992, A\&A, 253, 277.

\bibitem[Cliver et al. 1998]{cliv98}
Cliver, E.W., Boriakoff, V. \& Bounar, K.H. 1998, Geophys. Res. Lett.,
 25, 897


\bibitem[Eddy 1977a]{eddy77}
Eddy, J.A. 1977a, Sci. Am., 236(5), 80

\bibitem[Eddy 1977b]{eddy77a}
Eddy, J.A. 1977b, Clim. Change, 1, 173

\bibitem[Freeman et al. 2000]{free00}
Freeman, M.P., Watkins, N.W. \& Riley, D.J. 2006, Phys. Rev. E,
 62, 8794

\bibitem[Fr\"ohlich 2006]{froh06}
Fr\"ohlich, C., 2006, Space Sci. Rev., 125, 53.

\bibitem[Gleissberg 1944]{glei44}
Gleissberg, W. 1944, Terr. Magnet. Atmosph. Electr., 49, 243-244

\bibitem[Goslar 2003]{gosl03}
Goslar T. 2003, in: PAGES News (Past Global Changes), 11(2-3), 12

\bibitem[Hoyt \& Schatten 1998]{hoyt98}
Hoyt D.V. \& Schatten, K. 1998, Solar Phys. 179, 189

\bibitem[Korte \& Constable 2005]{kort05}
Korte, M. \& Constable, C.G. 2005, Earth Planet. Sci. Lett.,
 236, 348

\bibitem[Krivova, Balmaceda \& Solanki 2007]{balm07}
Krivova, N.A., Balmaceda, L. \& Solanki, S.K. 2007, A\&A, 467, 335.

\bibitem[Lepreti et al. 2001]{lepr01}
Lepreti, F., Carbone, V. \& Veltri P. 2001,
 Astrophys. J., 555, L133

\bibitem[Masarik \& Beer 1999]{masa99}
Masarik, J., \& J. Beer 1999, J. Geophys. Res., 104(D10), 12099.

\bibitem[McCracken et al. 2004]{mccr04}
McCracken, K.G., McDonald, F.B., Beer, J. et al. 2004,
 J. Geophys. Res., 109, A12103

\bibitem[Mega et al. 2003]{mega03}
Mega, M.S., Allegrini, P., Grigolini, P. et al. 2003,
 Phys. Rev. Lett., 90, 188501

\bibitem[Minini et al. 2001]{mini01}
Mininni, P.D., Gomez, D.O. \& Mindlin, G.B. 2001,
 Solar Phys., 201, 203

\bibitem[Miyahara et al. 2006]{miya06}
Miyahara, H., Sokoloff, D. \& Usoskin, I.G. 2006,
 The solar cycle at the Maunder minimum epoch,
 in: {\it Advances in Geosciences}, (eds. W.-H. Ip, M. Duldig)
 World Scientific, Singapore, pp.1-20.

\bibitem[Moon et al. 2001]{moon01}
Moon, Y.-J., Choe, G.S., Yun, H.S. \& Park, Y.D. 2001,
 J. Geophys. Res., 106, 29951

\bibitem[Mursula \& Ulich 1998]{murs98}
Mursula, K., \& Ulich, Th., 1998, Geophys. Res. Lett., 25, 1837.

\bibitem[Ossendrijver 2000]{osse00}
Ossendrijver, M. A. J. H., 2000, A\&A, 359, 1205.

\bibitem[Papa et al. 2006]{papa06}
Papa, A.R.R., Barreto, L.M. \& Seixas, N.A.B. 2006,
J. Atmos. Solar-Terr. Phys., 68, 930

\bibitem[Pearce et al. 1993]{pear93}
Pearce, G., Rowe, A.K., \& Yeung, J. 1993,
 Astrophys. Space Sci., 208, 99

\bibitem[Ponte-Neto \& Papa 2006]{pont06}
Ponte-Neto, C.F. \& Papa, A.R.R. 2006,
 eprint arXiv:physics/0602122

\bibitem[Schmitt et al. 1996]{schm96}
Schmitt, D., Sch\"ussler, M. \& Ferriz-Mas, A. 1996, A\&A, 311, L1.

\bibitem[Sokoloff 2004]{soko04}
Sokoloff, D.D. 2004, Solar Phys., 224, 145.


\bibitem[Solanki et al. 2000]{sola00}
Solanki, S. K., Sch\"ussler, M. \& Fligge, M. 2000, Nature, 408, 445.

\bibitem[Solanki et al. 2002]{sola02}
Solanki, S. K., Sch\"ussler, M. \& Fligge, M. 2002, A\&A, 383, 706.

\bibitem[Solanki et al. 2004]{sola04}
Solanki, S.K., Usoskin, I.G., Kromer, B., Sch\"ussler, M. \& Beer, J. 2004,
 Nature, 431, 1084

\bibitem[Soon, Posmentier \& Baliunas 1996]{soon96}
Soon, W. H., Posmentier, E. S., \& Baliunas, S. L., 1996, ApJ, 472, 891.

\bibitem[Stuiver 1980]{stui80}
Stuiver, M. 1980, Nature, 286, 868

\bibitem[Stuiver \& Braziunas 1989]{stui89}
Stuiver, M. \& Braziunas T.F. 1989, Nature, 338, 405.

\bibitem[Stuiver et al. 1991]{stui91}
Stuiver, M., Braziunas T.F., Becker B. \& Kromer, B. 1991, Quatern. Res., 35, 1.

\bibitem[Stuiver et al. 1998]{stui98}
Stuiver, M., Reimer, P.J., Bard, E. et al. 1998, Radiocarbon, 40, 1041

\bibitem[Usoskin et al. 2001]{usos01}
Usoskin, I.G., Mursula, K. \& Kovaltsov, G.A. 2001
 J. Geophys. Res., 106, 16039

\bibitem[Usoskin et al. 2002]{usos02}
Usoskin, I.G., K. Mursula, S. Solanki, Sch\"ussler, M. \& Kovaltsov, G.A. 2002,
 J. Geophys. Res., 107(A11), 1374.

\bibitem[Usoskin et al. 2003]{usos03}
Usoskin, I.G., Solanki, S.K., Sch\"ussler, M., Mursula, K. \& Alanko, K. 2003,
 Phys. Rev. Lett., 91, 211101

\bibitem[Usoskin et al. 2004]{usos04}
Usoskin, I.G., K. Mursula, S. Solanki, Sch\"ussler, M. \& Alanko, K. 2004, A\&A, 413, 745.

\bibitem[Usoskin et al. 2005]{usos05}
Usoskin, I.G., K. Alanko-Huotari, G.A. Kovaltsov \& K. Mursula, 2005,
 J. Geophys. Res., 110, A12108.

\bibitem[Usoskin et al. 2006a]{usos06}
Usoskin, I.G., Solanki, S.K., \& Korte, M. 2006a,
 Geophys. Res. Lett., 33, L08103.

\bibitem[Usoskin et al. 2006b]{usos06b}
Usoskin, I.G., S.K. Solanki, C. Taricco, N. Bhandari \& G.A. Kovaltsov, 2006b,
 A\&A, 457, L25.

\bibitem[Vasiliev \& Dergachev 2002]{vasi02}
Vasiliev, S.S., \& V.A. Dergachev, 2002, Annales Geophys., 20, 115.

\bibitem[Voss et al. 1996]{voss96}
Voss, H., Kurths, J. \& U. Schwarz, 1996, J. Geophys. Res., 101, 15637

\bibitem[Weiss \& Tobias 2000]{weis00}
Weiss, N.O. \& Tobias, S.M., 2000, Space Sci. Rev., 94, 99.

\bibitem[Wheatland 2003]{whea03}
Wheatland, M.S. 2003, Solar Phys., 214, 361

\bibitem[Yang et al. 2000]{yang00}
Yang, S., Odah, H. \& Shaw, J. 2000, Geophys. J. Int., 140, 158

\end{thebibliography}
\end{document}